\documentclass[useAMS,usenatbib]{mnras}
\usepackage[english]{babel}
\usepackage[fleqn]{amsmath}
\usepackage{amsfonts}
\usepackage{txfonts}
\usepackage{graphicx}
\usepackage{amssymb}
\usepackage[T1]{fontenc}
\usepackage{ae,aecompl}
\usepackage{fancyhdr}
\usepackage{multicol}
\usepackage{layout}
\usepackage{graphicx}
\usepackage{multirow}
\usepackage{color}
\usepackage{multirow}

\usepackage{times}

\usepackage[outdir=./]{epstopdf}

\newif\ifAMStwofonts
\AMStwofontstrue

\def\rev#1{{#1}}

\title[Spherical collapse in tidal fields]{Shear and vorticity in the spherical collapse of dark matter haloes}
\author[R. Reischke, F. Pace, S. Meyer, B. M. Sch{\"a}fer]{Robert Reischke$^{1}$\thanks{e-mail:
reischke@stud.uni-heidelberg.de}, Francesco Pace$^2$, Sven Meyer$^3$ and Bj\"orn Malte Sch\"afer$^{1}$\\
$^{1}$ Zentrum f{\"u}r Astronomie der Universit{\"a}t Heidelberg, Astronomisches Recheninstitut, Philosophenweg 12, 
69120 Heidelberg, Germany\\
$^{2}$ Jodrell Bank Centre for Astrophysics, School of Physics and Astronomy, The University of Manchester, Manchester, 
M13 9PL, United Kingdom\\
$^{3}$Zentrum f{\"u}r Astronomie der Universit{\"a}t Heidelberg, Institut f{\"u}r theoretische Astrophysik, 
Philosophenweg 12, D-69120, Heidelberg, Germany}
\begin{document}

\date{2017}

\pagerange{\pageref{firstpage}--\pageref{lastpage}} \pubyear{}
\maketitle

\label{firstpage}

\begin{abstract}
{\rev{ 
Traditionally the spherical collapse of objects is studied with respect to a uniform background density, yielding the 
critical over-density $\delta_\mathrm{c}$ as key ingredient to the mass function of virialized objects. Here we 
investigate the shear and rotation acting on a peak in a Gaussian random field. By assuming that collapsing objects 
mainly form at those peaks, we use this shear and rotation as external effects changing the dynamics of the spherical 
collapse, which is described by the Raychaudhuri equation. We therefore assume that the shear and rotation have no 
additional dynamics on top of their cosmological evolution and thus only appear as inhomogeneities in the differential 
equation.}}

{\rev{
We find that the shear will always be larger than the rotation at peaks of the random field, which automatically 
results into a lower critical over-density $\delta_\mathrm{c}$, since the shear always supports the collapse, while the 
rotation acts against it. Within this model $\delta_\mathrm{c}$ naturally inherits a mass dependence from the Gaussian 
random field, since smaller objects are exposed to more modes of the field.
The overall effect on $\delta_\mathrm{c}$ is approximately of the order of a few percent with a decreasing trend to 
high masses.
}}
\end{abstract}


\section{Introduction}
The combined observations of type-Ia supernovae \citep[e.g.][]{Riess1998,Perlmutter1999}, the cosmic microwave 
background \citep[e.g.][]{Komatsu2011,Planck2016_XIII}, the Hubble constant and large-scale structure 
\citep[e.g.][]{Cole2005} show that the universe is spatially flat and expanding in an accelerated fashion. 
Under the assumptions of General Relativity being true and the symmetries of the Friedmann-Robertson-Walker metric, 
the accelerated expansion can be described by the cosmological constant $\Lambda$ or by adding a fluid component 
\citep[see e.g.][for a review]{Copeland2006} to the energy-momentum content of the Universe with an equation of state 
$w<-1/3$ which may be time dependent. In this scenario, generally called dark energy, $\Lambda$ would corresponds to a 
constant $w=-1$.

The equation of state parameter, as well as the other parameters such as the dark energy content $\Omega_{\Lambda}$ or 
the matter content $\Omega_\mathrm{m}$ of the cosmological standard model can and already have been measured to very 
high precision. However, with high precision comes also the risk of possible biases in the parameters if the 
theoretical prediction and astrophysics are not modelled well enough. 
It is therefore necessary to review and perhaps modify common concepts and models since possible systematics are no 
longer wiped out by the statistical errors of the experiment.

The halo mass function is, due to its exponential sensitivity, one of the main tools to provide robust theoretical 
predictions for many observables such as cluster counts 
\citep{Sunyaev1980b,Majumdar2004,Diego2004,Fang2007,Abramo2009a,Angrick2009} or weak lensing peak counts 
\citep{Maturi2010,Maturi2011,Lin2014,Reischke2016}. As it deals with objects in the highly non-linear regime one needs 
to extrapolate the linearly evolved density to the non-linear one which is encoded in the critical linear overdensity 
$\delta_\mathrm{c}$. 

This can be done by using the spherical collapse model introduced by \citet{Gunn1972} and later 
extended in several works 
\citep{Fillmore1984,Bertschinger1985,Ryden1987,AvilaReese1998,Mota2004,Abramo2007,Pace2010,Pace2014}. 
This model assumes in its most simplistic form, called standard spherical collapse (SPC hereafter), the evolution of a 
spherically symmetric density perturbation in an expanding background with uniform background density, i.e. an isolated 
collapse. The overdensity grows until it reaches a critical point at which it starts to collapse under its own gravity. 
Due to the collapse it decouples from the expansion. In theory the overdense region would collapse to a single point, 
however, the energy released during the collapse is converted into random motions of the particles, such that an 
equilibrium situation \citep[in the sense of virialized structure,][]{Schafer2008} is created. {\rev{Effectively we start from initial conditions such that such a virialized structure is formed at a certain redshift. These initial conditions are then used to propagate the linear growth of structures, described by the overdensity $\delta$. The corresponding linear $\delta$ at the redshift where the non-linear solution resulted into a virialized object is then called $\delta_\mathrm{c}$ It therefore sets a threshold above which an object can be considered as being collapsed even in terms of linear structure formation. Given this value one can predict the abundances of collapsed objects in terms of a random walk with a moving barrier.}}

It is necessary to check the validity of the SPC model and introduce additional effects subject to more realistic 
situations. One of the ingredients of the SPC model is the embedding of the spherical region into a uniform background. 
In this way no external forces can act on the halo by virtue of Birkhoff's theorem. {\rev{However, halos are embedded 
in a random field giving rise to gravitational tidal fields inducing shear and rotation acting on the collapsing 
spherical region and thus changing the collapse dynamics.}}

In earlier works \citep{DelPopolo2013a,DelPopolo2013b,Pace2014b}, the influence of rotation and shear has been 
studied extensively using a phenomenological model introducing an additional term to match the Newtonian predictions. 
The authors found that the effect of shear is always smaller than the effect of the rotation, leading to a slowdown of 
the collapse. Consequently $\delta_\mathrm{c}$ is larger compared to the SPC scenario, where it adopts the known value 
1.686. The additional term has a mass dependence leading to a mass-dependent $\delta_\mathrm{c}$ as well. Lower mass 
halos are exposed to higher values of rotation and shear, leading to a higher $\delta_\mathrm{c}$ while it becomes 
negligible for halos with $M\gtrsim 10^{15}M_\odot$.

More recently \citet{Reischke2016a} (R16 from now on) investigated a test particle embedded in tidal gravitational 
fields described by a Gaussian random field. The Gaussian random field obeys the statistics of the cosmic density field 
$\delta$ by virtue of the Zel'Dovich approximation \citep{zeldovich1970}. This gives rise to an effective shear acting 
on the collapsing region. The mass dependence of the shear is introduced naturally by assigning a length scale to an 
object of mass $M$. Since the shear is treated as a random variable, $\delta_\mathrm{c}$ is a random variable as well 
and has a distribution rather than a distinct value and the averaged value should be used for the mass function. It was 
shown in R16 that the effect on $\delta_\mathrm{c}$ can cause a $1\sigma$ bias in cosmological parameters when 
considering an idealized cluster survey. \citet{Pace2017} studied the effect of shear in clustering dark energy models, 
finding similar results to the smooth case.

Since the tidal field examined in R16 is described by a potential flow there is no vorticity generation. However, 
a rotation of the collapsing region can be modelled by a mechanism called tidal torquing 
\citep{White1984,Catelan1996a,Crittenden2001,Schafer2009,Schafer2012}. 
We will therefore consider a peak in the density field with inertial tensor $\boldsymbol{I}$ and tidal shear tensor 
$\boldsymbol{\Psi}$ and investigate jointly the induced shear and rotation. {\rev{Assuming that halos form}} at peaks, 
we will use the values estimated for the shear and the rotation as input for the spherical collapse model leading to a 
self-consistent description of the spherical collapse in gravitational tidal fields. We will furthermore show that the 
restriction to peaks in the density field has some very general consequences on the induced rotation and shear.

The structure of the paper is as follows. In \autoref{sec:SC} we very briefly review the spherical collapse model and 
show the equations to be solved. In \autoref{sec:RotandShear} we introduce the statistical procedure to obtain tidal 
shear values and decompose them to identify the shear and rotation tensor respectively.  
The obtained invariants are then used in \autoref{sec:deltac} to calculate the influence of the tidal fields on 
$\delta_\text{c}$ for the standard $\Lambda$CDM model. 
We summarize our findings in \autoref{sec:concl}.

\section{Spherical Collapse}\label{sec:SC}
The spherical collapse model has been discussed by various authors, e.g. 
\citet{Bernardeau1994,Padmanabhan1996,Ohta2003,Ohta2004,Abramo2007} and \citet{Pace2010,Pace2014} {\rev{and in its 
standard scenario describes the evolution of a spherical over-dense region in an homogeneous expanding background. The 
key quantity derived from the spherical collapse is the critical over-density $\delta_\mathrm{c}$ which allows an 
extrapolation from the linear evolved density field to virialized structures. Therefore $\delta_\mathrm{c}$ is primal 
for the study of the abundance of objects in the Universe. }}
Starting from the perturbed hydrodynamical equations
\begin{equation}\label{eq:1}
 \begin{split}
  \dot{\delta} + (1+\delta)\nabla_{\boldsymbol x}\boldsymbol u = & \ 0\;, \\ 
  \dot{\boldsymbol u} + 2H\boldsymbol u+ (\boldsymbol u\cdot\nabla_{\boldsymbol x})\boldsymbol u = & \ 
  -\frac{1}{a^2}\nabla_{\boldsymbol x}\phi \;, \\
  \nabla^2_{\boldsymbol x}\phi = & \ 4\pi G a^2\rho_0 \delta\;,
 \end{split}
\end{equation}
with comoving coordinate $\boldsymbol x$, comoving peculiar velocity $\boldsymbol u$, Newtonian potential $\phi$, 
overdensity $\delta$ and background density $\rho_0$, respectively. Here the dot represents a derivative with respect 
to cosmic time $t$. Taking the divergence of the Euler equation and inserting the Poisson equation yields
\begin{equation}\label{eq:2}
 \begin{split}
  \dot\delta = & -(1+\delta)\theta\;, \\
  \dot\theta = & -2H\theta -4\pi G\rho_0 \delta-\frac{1}{3}\theta^2 -(\sigma^2-\omega^2)\;.
 \end{split}
\end{equation}
Here we used the decomposition
\begin{equation}\label{eq:3}
 \nabla_{\boldsymbol x} \cdot [(\boldsymbol u\nabla_{\boldsymbol x}) \boldsymbol u)] = 
 \frac{1}{3}\theta^2 +\sigma^2-\omega^2\;,
\end{equation}
with the expansion $\theta = \nabla_{\boldsymbol x}\cdot\boldsymbol u$, the shear 
$\sigma^2 \equiv \sigma_{ij}\sigma^{ij}$ and the rotation $\omega^2\equiv \omega_{ij}\omega^{ij}$ {\rev{assuming 
spherical symmetry}}. The rotation and the shear tensors are themselves the anti-symmetric and the symmetric traceless 
part of the velocity divergence tensor, respectively. They are defined as
\begin{equation}
 \begin{split}
  \sigma_{ij} = & \ \frac{1}{2}\left(\partial_i u_j +\partial_j u_i\right) - \frac{\theta}{3}\delta_{ij}\;, \\
  \omega_{ij} = &  \ {\frac{1}{2}}\left(\partial_i u_j -\partial_j u_i\right)\;,
 \end{split}
\end{equation}
where $\partial_i\equiv \partial/\partial x^i$. 
We now use the relation $\partial_t = aH(a)\partial_a$ and $\tilde{\delta}=1/\delta$ which leads to
\begin{equation}\label{eq:4}
 \begin{split}
 \tilde{\delta}^{\prime} = & \ \frac{\theta}{aH}\tilde{\delta}(1+\tilde{\delta})\;, \\
 \theta^{\prime} = & -\frac{2\theta}{a}-\frac{3H\Omega_{\text{m}}}{2a\tilde{\delta}}-
 \left(\frac{1}{3}\theta^2+\sigma^2-\omega^2\right)\frac{1}{aH}\;,
 \end{split}
\end{equation}
where the prime denotes a derivative with respect to $a$.
The system defined in Eq.~(\ref{eq:4}) is solved numerically until $\tilde{\delta}\sim 10^{-8}$ and then it is 
extrapolated to zero, which is much more stable than treating the system in $\delta$ rather than $\tilde{\delta}$. 
This yields the appropriate initial conditions for the linear version of \ref{eq:4} which gives $\delta_{\rm c}$. 
Usually $\sigma^2$ and $\omega^2$ are neglected, however, their influence has phenomenologically been investigated, 
e.g. by \citet{DelPopolo2013a,DelPopolo2013b} in the $\Lambda$CDM and dark energy cosmologies and by \cite{Pace2014b} 
in clustering dark energy models. 
The authors heuristically model the term $\sigma^2-\omega^2$ allowing to study an isolated collapse including a 
(mass dependent) quantity $\alpha$, defined as the ratio between the rotational and the gravitational term. 
Specifically, the term is
\begin{equation}
 \alpha=\frac{L^2}{M^3RG}\;,
\end{equation}
where $L$ is the angular momentum of the spherical overdensity considered. $M$ and $R$ are its mass and radius, 
respectively. The angular momentum is important for galaxies, but negligible for clusters. In particular 
$\alpha\approx 0.05$ for $M\approx 10^{11}~{\rm M}_{\odot}~h^{-1}$ and of the order of $10^{-6}$ for 
$M\approx 10^{15}~{\rm M}_{\odot}~h^{-1}$. By defining new quantities $\tilde{\theta}=\theta/H$, 
$\tilde{\sigma}=\sigma/H$ and $\tilde{\omega}=\omega/H$, the combined contribution of the shear and rotation term can 
effectively be modelled by
\begin{equation}
 \tilde{\sigma}^2-\tilde{\omega}^2=-\frac{3}{2}\alpha\Omega_{\mathrm{m}}\delta\;,
\end{equation}
leading to modified equations for the spherical collapse. Clearly $\tilde{\sigma}^2-\tilde{\omega}^2$ is always 
negative, thus leading to a slowdown of the collapse. Technically the terms leads to larger initial conditions for the 
density contrast for a halo to be formed at a certain redshift, thus increasing $\delta_\mathrm{c}$ as it is obtained 
from the linear equation using these initial conditions.

\begin{figure}
 \begin{center}
  \includegraphics[width = 0.45\textwidth]{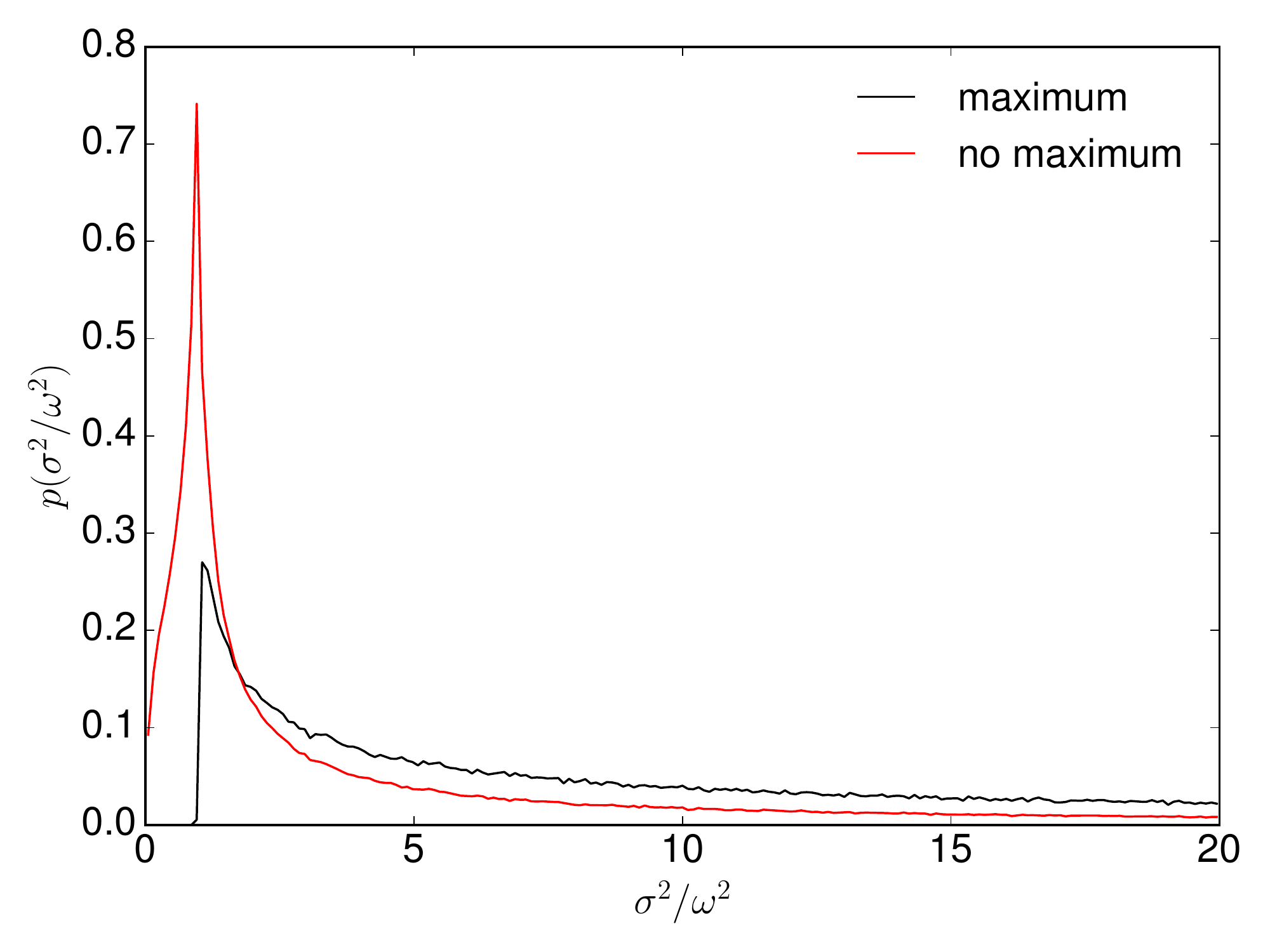}
  \caption{Distributions of the ratio of the two invariants $\sigma^2$ and $\omega^2$. The {\bf red} histogram does not 
  include the maximum constraint, i.e. $\lambda_i > 0$, while the {\bf black} histogram includes this constraint. 
  Clearly the constraint moves all values which would have $\sigma^2<\omega^2$ to values $\sigma^2>\omega^2$ as it is 
  expected from the analytical considerations made in Eq.~(\ref{eq:positivedefinit}). The smoothing length for the 
  power spectrum is $R=10\,\text{Mpc}h^{-1}$.}
  \label{Fig:sigmaomega}
 \end{center}
\end{figure}

\begin{figure*}
 \begin{center}
  \includegraphics[width = 0.45\textwidth]{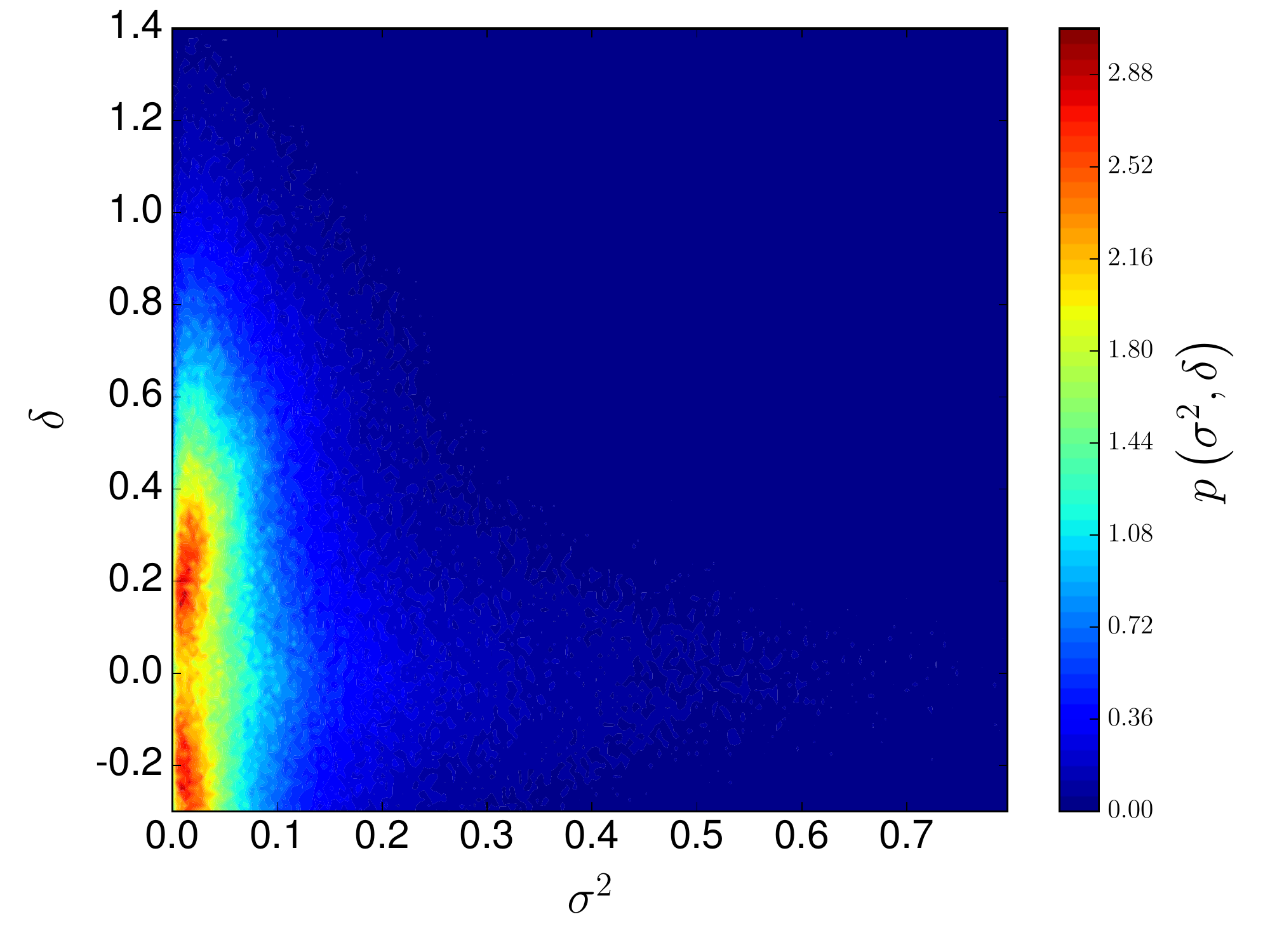}
  \includegraphics[width = 0.45\textwidth]{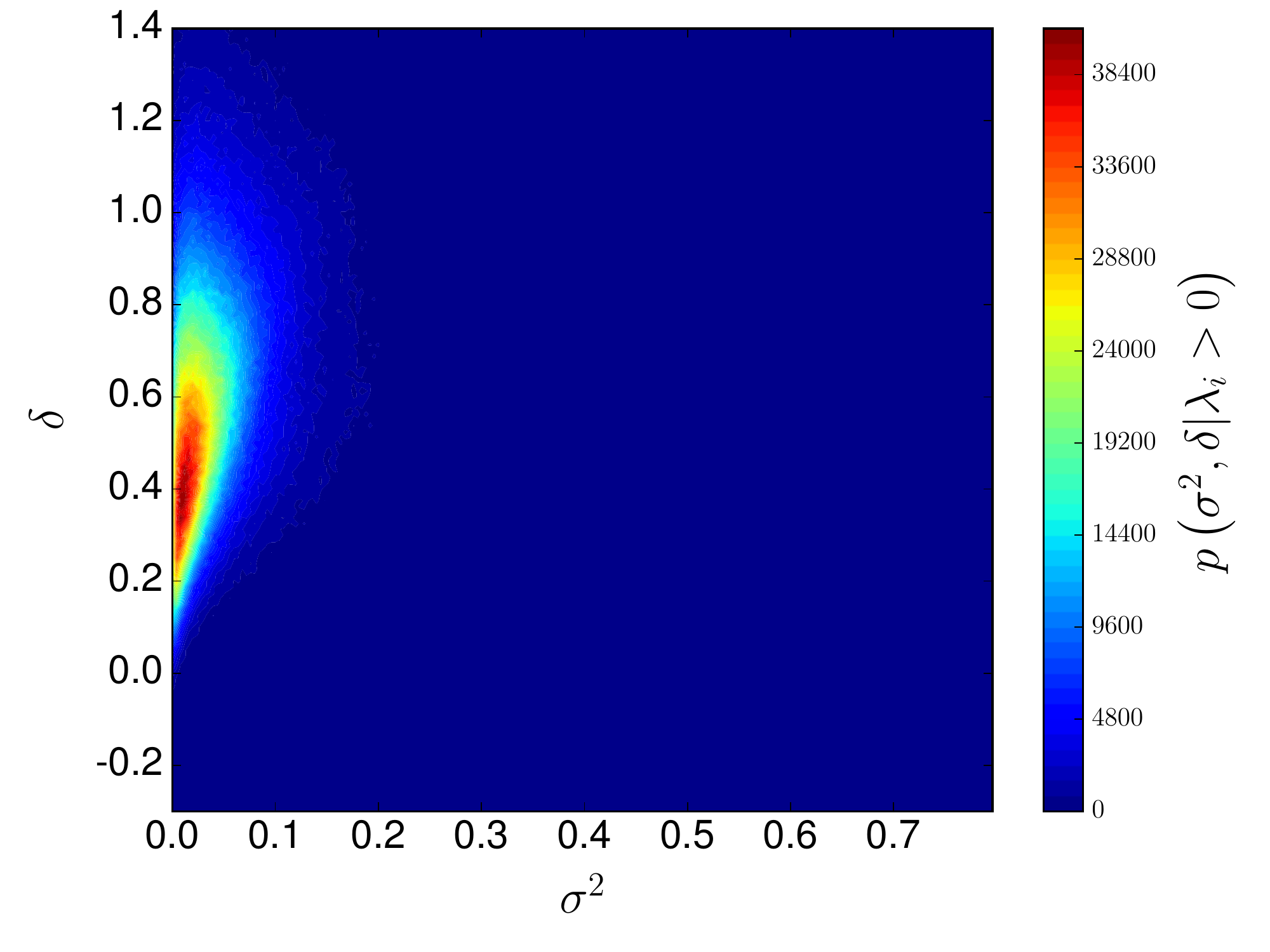}
  \caption{Joint distributions (not normalized) of the density contrast $\delta = \text{tr}(\boldsymbol{\Psi})$ and the 
  invariant $\sigma^2$. 
  \textit{Left panel}: No maximum constraint. The distribution is similar to the distribution found in R16. 
  \textit{Right panel}: The maximum constraint is imposed. Clearly the constraint enforces higher values of $\delta$ 
  and thus also in $\sigma^2$, which is due to the correlations in the $y^n_{lm}$ basis given in 
  Eq.~(\ref{eq:correlations}). The smoothing length for the power spectrum is $R=10\,\text{Mpc}h^{-1}$.}
  \label{Fig:Correlations}
 \end{center}
\end{figure*}

\section{Shear and Rotation}\label{sec:RotandShear}

\subsection{The model}
{\rev{In this work we intend to model the shear invariant $\sigma^2$ together with the rotation invariant $\omega^2$ 
which occur in the collapse equation (\ref{eq:4}). Earlier works studied the joint influence of shear and rotation in a 
phenomenological way as mentioned in the last section. The logic here is the following: We assume that dark matter 
halos form at peaks of the density field, which itself is described by a Gaussian random field and thus by its power 
spectrum. In order to model $\sigma^2 -\omega^2$ we calculate these values at peaks of the density field using only 
the statistics of the field itself and the Zel'Dovich approximation.
We then effectively place a test particle into the Gaussian random field at the peak and let it undergo gravitational collapse, with the shear and rotation acting as external forces with no own dynamics (except for the ones given by the background dynamics). Thus the collapse dynamics will stay spherical, 
while we allow for deviations from sphericity in the estimation of the shear and rotation (especially to find an 
expression for the inertial tensor). In this way the collapsing object can be seen as a test particle in a tidal 
gravitational field.}}

\subsection{The tidal tensor}
The central object of our model is the tidal tensor $\boldsymbol{\Psi}$, which is related to the density field.
For scales large enough, particles follow Zel'Dovich trajectories \citep{zeldovich1970}
\begin{equation}\label{eq:Zeld}
 x_i = q_i-D_+(t)\partial_i \psi \equiv q_i -D_+(t)\psi_{,i}\;.
\end{equation}
The displacement field $\psi$ is related to the density contrast $\delta$ via a Poisson relation, 
$\Delta \psi = \delta$. In Fourier space we can thus write the components of the tidal tensor as:
\begin{equation}
 \psi_{,ij} = -\int\frac{\text{d}^3k}{(2\pi)^3}\frac{k_ik_j}{k^2}\delta(\boldsymbol k)\exp(\text{i}\boldsymbol{k}
               \boldsymbol{x})\;.
\end{equation}
We now choose spherical coordinates in such a way that the peaks under consideration lie symmetric around the origin on 
the $z$-axis \citep{Regos1995,Heavens1999}. For convenience we introduce dimensionless complex variables
\begin{equation}
 y^n_{lm} = \sqrt{4\pi}\frac{\text{i}^{l+2n}}{\sigma_{l+2n}}\int\frac{\text{d}^3k}{(2\pi)^3}k^{l+2n}
 \delta(\boldsymbol k) Y_{lm}(\hat{k})\exp(\text{i}\boldsymbol k\boldsymbol x)\;,
\end{equation}
with the direction vector $\hat{k}=\boldsymbol k/k$ and $\sigma_i$ being the spectral moments of the matter power 
spectrum
\begin{equation}\label{eq:spectralmoments}
 \sigma_i^2 = \frac{1}{2\pi^2}\int\text{d}k\; k^{2i+2} P(k)\;.
\end{equation}
and $Y_{lm}$ are the spherical harmonics. With this we obtain a linear relation \citep{Schafer2012} between $y^n_{lm}$ 
and the tidal field values $\psi_{,ij}$
\begin{equation}\label{eq:11}
 \begin{split}
  \sigma_0 y_{20}^{-1} = & \ -\sqrt{\frac{5}{4}}\left(\psi_{,xx}+\psi_{,yy}-2\psi_{,zz}\right)\;, \\
  \sigma_0 y^{-1}_{2\pm 1} = & \ -\sqrt{\frac{15}{2}}\left(\psi_{,xz}\pm\text{i}\psi_{,yz}\right)\;, \\
  \sigma_0 y^{-1}_{2\pm 2} = & \ \sqrt{\frac{15}{8}}\left(\psi_{,xx}-\psi_{,yy}\pm2\text{i}\psi_{,xy}\right)\;, \\
  \sigma_0 y^{0}_{00} = & \ \left(\psi_{,xx}+ \psi_{,yy}+\psi_{,zz}\right)\;.
 \end{split}
\end{equation}
The variables $y^n_{lm}$ now have a diagonal auto-correlation matrix:
\begin{equation}\label{eq:correlations}
 \left\langle y_{lm}^n(\boldsymbol x)y_{l^{\prime}m^{\prime}}^{n^{\prime}}(\boldsymbol x)^{\ast}\right\rangle = 
 (-1)^{n-n^{\prime}}\frac{\sigma^2_{l+n+n^{\prime}}}{\sigma_{l+2n}\sigma_{l+2n^{\prime}}}
 \delta_{ll^{\prime}}\delta_{mm^{\prime}}\;.
\end{equation}
By inverting the linear relation, we can draw random samples in the $y^n_{lm}$ basis and calculate the components of 
the tidal tensor $\boldsymbol{\Psi}_{ij} \equiv \psi_{,ij}$. 
The strength of the tidal field depends on the characteristic length scale $R(M)$ of the halo and thus on its mass. We 
thus introduce a low-pass filter to cut off high frequency modes, suppressing fluctuations on scales smaller than the 
characteristic scale of the halo:
\begin{equation}
 P(k) \to P(k)W^2_R(k)\;, 
\end{equation}
with $W_R(k) = \exp(-k^2R^2/2)$. The mass scale is obtained via 
$M = \frac{4\pi}{3}\rho_\text{crit}\Omega_\text{m} R^3$, where $\rho_\text{crit}= 3H^2/(8\pi G)$ is the critical 
density. For more details on the tidal field and the random process we refer to R16.

\subsection{Tidal Torquing}  
{\rev{Having presented the procedure to sample the components of the tidal tensor directly from the statistics of the 
density field, we introduce a mechanism known as tidal torquing in order to describe the generation of rotation due to 
tidal gravitational fields.}}
In the picture of tidal torquing angular momentum is generated by the tidal gravitational field which exerts a torquing 
moment on the halo. It is important to note that the vorticity $\boldsymbol{\omega}$ is not driven by the 
non-linear term $\nabla\times(\boldsymbol{v}\times\boldsymbol{\omega})$ in the Euler equation. On the contrary, the 
angular momentum is generated by vorticity-free flows generating shear effects on the halo prior to collapse. 
During this process the halo is slightly deformed and tends to align its inertia tensor in the eigenframe of the shear 
tensor. After decoupling from the shear flow and the start of collapse the length of the lever arms reduces 
dramatically {\rev{in comoving coordinates}} making tidal torquing inefficient. Therefore, the angular momentum just 
before collapse begins is a good proxy for the total rotation of the halo. 

\begin{figure}
 \begin{center}
  \includegraphics[width = 0.45\textwidth]{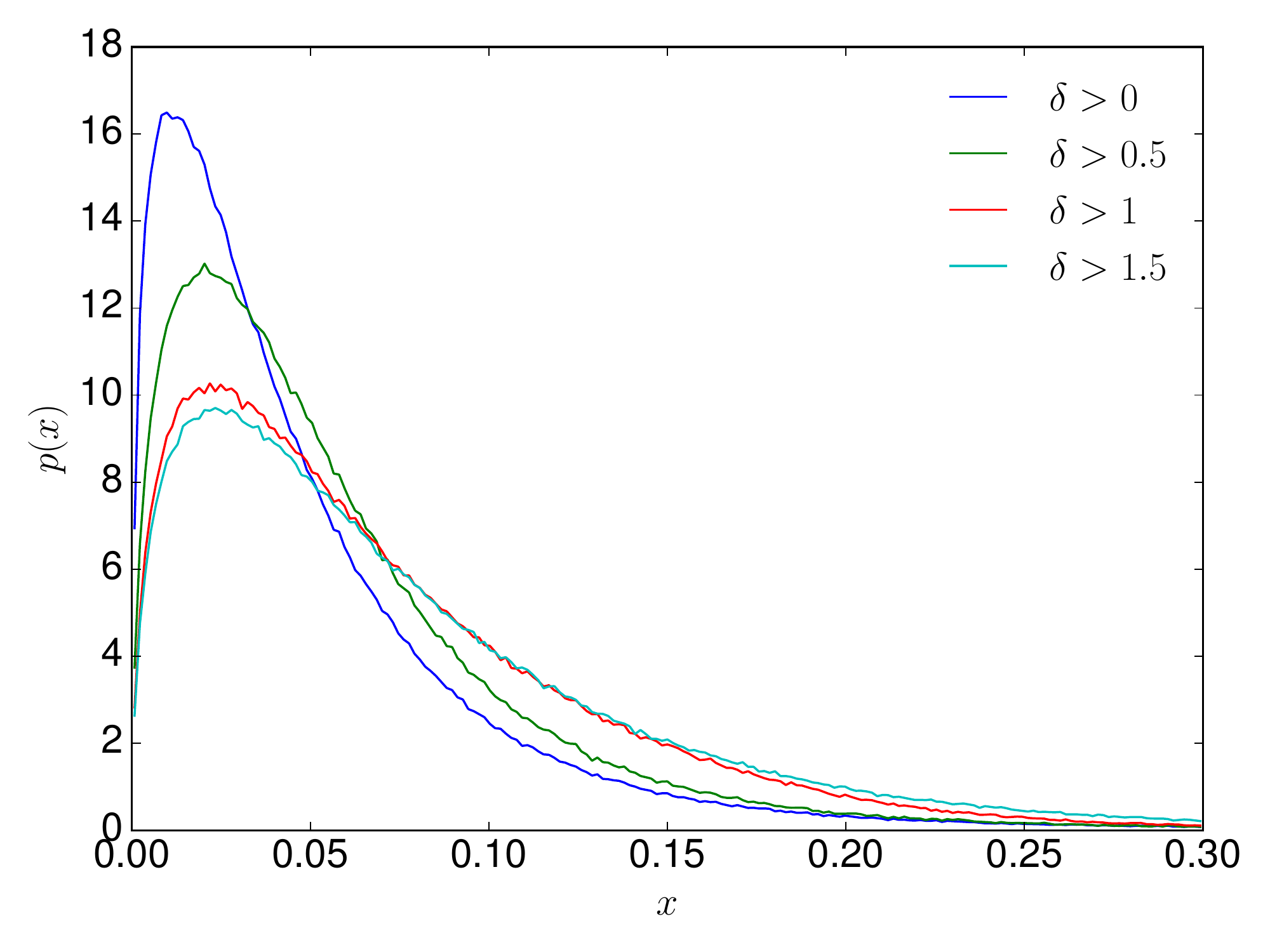}
  \caption{Distributions of the invariants $\sigma^2 - \omega^2$ for different thresholds $\delta$ on a 
  scale of $R=10\,\text{Mpc}h^{-1}$. Higher peaks induce higher values for $\sigma^2 - \omega^2$.}
  \label{Fig:Compareprevious}
 \end{center}
\end{figure}

The angular momentum $\boldsymbol{L}$ of a rotating mass distribution $\rho(\boldsymbol{r},t)$ is given by
\begin{equation}
 \boldsymbol{L}(t) = \int_V\text{d}^3r\:(\boldsymbol{r}-\boldsymbol{\bar r})\times 
                     \boldsymbol{v}(\boldsymbol{r},t)\rho(\boldsymbol{r},t)\;,
\end{equation}
with $\boldsymbol{v}$ being the rotational velocity and $V$ the physical volume under consideration. Making use of the 
Zel'Dovich approximation and expressing everything in the Lagrangian-frame (i.e. comoving), the angular momentum 
becomes
\begin{equation}
 \boldsymbol{L} = \rho_0 a^5\int_{V_L}\text{d}^3q\:(\boldsymbol{q}-\boldsymbol{\bar q})\times \boldsymbol{\dot x}\;,
\end{equation}
neglecting higher order terms \citep{White1984,Catelan1996a,Crittenden2001}. The velocity $\boldsymbol{\dot x}$ is 
given via the gradient of the potential $\psi$, which can be expanded in the vicinity of the centre of gravity 
$\boldsymbol{\bar q}$ if its variation across the Lagrangian volume $V_L$ is small:
\begin{equation}
 \partial_i \psi(\boldsymbol{q}) \approx \partial_i\psi(\boldsymbol{q})\big|_{\boldsymbol{q}=\boldsymbol{\bar q}}+
 \partial _{ij}\psi (q)\big|_{\boldsymbol{q}=\boldsymbol{\bar q}}(\boldsymbol{q}-\boldsymbol{\bar q})_j\;,
\end{equation}
with expansion coefficients $\psi_{ij} \equiv \partial_{ij}\psi$ describing the tidal shear given in Eq.~(\ref{eq:11}). 
The first term can be neglected as it only describes the displacement of the protohalo, the second however will be 
responsible for the rotational effects. Introducing the inertial tensor $I_{ij}$ as
\begin{equation}
 I_{ij} = \rho_0a^3\int_{V_l}\text{d}^3q\:
          (\boldsymbol{q}-\boldsymbol{\bar q})_i(\boldsymbol{q}-\boldsymbol{\bar q})_j\;,
\end{equation}
the angular momentum can be written as
\begin{equation}
 L_i = a^2\dot D_+\epsilon_{ijk}I_{jl}\psi_{lk}\;,
\end{equation}
with the Levi-Civita symbol $\epsilon_{ijk}$. The matrix product in the latter expression 
$\boldsymbol{X}=\boldsymbol{I}\boldsymbol{\Psi}$ can be decomposed into a symmetric $\boldsymbol{X}^+$ and 
anti-symmetric part $\boldsymbol{X}^{-}$ defined via the anti-commutator and the commutator, respectively:
\begin{equation}
 \boldsymbol{X}^{+} \equiv \frac{1}{2}\{\boldsymbol{I},\boldsymbol{\Psi}\}\;, \quad 
 \boldsymbol{X}^{-} \equiv \frac{1}{2}[\boldsymbol{I},\boldsymbol{\Psi}]\;.
\end{equation}
With this definition the angular momentum can be written as \citep{Schafer2009,Schafer2012}
\begin{equation}
 L_{i} = a^2\dot D_+\epsilon_{ijk}X_{jk} = a^2\dot D_+\epsilon_{ijk}X^{-}_{jk},
\end{equation}
since the contraction with $\epsilon_{ijk}$ will only pick out the anti-symmetric part of $\boldsymbol{X}$. Thus, 
angular momentum is not generated if inertia and tidal shear have a common eigensystem, which is always the case for a 
matter distribution invariant under SO(3), therefore we need to have $\boldsymbol{X}^{-}\ne 0$ to generate angular 
momentum. On the other hand $\boldsymbol{X}^{+}$ will measure the alignment of the eigensystems of inertia and shear 
and thus cause shear effects due to deformations.

The components of the inertial tensor $\boldsymbol{I}$ can be expressed via second derivatives of the density field 
$\delta(\boldsymbol{x})$ which are given by
\begin{equation}
 \delta_{ij} = -\int \frac{\mathrm{d}^3k}{(2\pi)^3}k_ik_j\delta(\boldsymbol{k})
                    \exp{(\text{i}\boldsymbol{k}\boldsymbol{x})}\;.
\end{equation}
Thus the decomposition works in the same way as before:
\begin{equation}\label{eq:curvaturedens}
 \begin{split}
  \sigma_2 y_{20}^0 = & \  -\sqrt{5/4}\left(\delta_{xx}+\delta_{yy}-\delta_{zz}\right)\;, \\
  \sigma_2 y_{2\pm 1}^0 = &\ -\sqrt{15/2}\left(\delta_{xz} \pm\text{i}\delta_{yz}\right)\;, \\
  \sigma_2y_{2\pm 2}^0 = & \ \sqrt{15/8}\left(\delta_{xx} -\delta_{yy}\pm 2\text{i}\delta_{xy}\right)\;, \\
  \sigma_2y_{00}^1 = & \ (\delta_{xx} +\delta_{yy} + \delta_{zz})\;.
 \end{split} 
\end{equation}
At a peak in the density field, the peak slope is approximated by a parabolic function
\begin{equation}
 \delta(\boldsymbol{x}) = \boldsymbol{x}_p -\frac{1}{2}\lambda_i(\boldsymbol{x}-\boldsymbol{x}_p)_i^2\;,
\end{equation}
with the eigenvalues $\lambda_i$ of the mass tensor $m_{ij} = -\delta_{ij}$ at the peak. If the boundary of the peak is 
given by the isodensity contour with $\delta = 0$, the inertia tensor can be written as
\begin{equation}\label{eq:inertialtensor}
 \boldsymbol{I} = \frac{\eta_0}{5}\Gamma~\text{diag}\left(A_y^2+A_z^2,A_x^2+A_z^2, A_x^2+A_y^2\right)\;,
\end{equation}
{\rev{in the eigen-system of the paraboloid.}} 
Here $A_i=\sqrt{2\delta/\lambda_i}$ are the ellipsoids semi-axes, $\Gamma$ its volume and $\eta_0$ its density, such 
that $M = \eta_0\Gamma$ is the mass of the peak. In our approximation the density field is assumed to be homogeneous to 
first order and thus $\eta_0 =\Omega_\text{m}\rho_\text{crit}a^3$. We thus sample values for $\boldsymbol{X}^{\pm}$ 
from the joint covariance matrix of $\delta_{ij}$ and $\psi_{ij}$. 
{\rev{All calculations are carried out in the eigen-system of the inertia tensor, i.e. we sample values $y_{lm}^n$ and 
calculate the inertia tensor by inverting Eq. (\ref{eq:curvaturedens}) and using Eq. (\ref{eq:inertialtensor})}}.

\subsection{Decomposition of the shear tensor}
{\rev{In the last two parts we described how the statistics of the density field induce tidal gravitational fields, 
encoded in $\boldsymbol{\Psi}$, and how these tidal fields can give rise to rotation. Since the shear effects are as 
well described by the tidal tensor the scope of this section will be to decompose $\boldsymbol{\Psi}$ into two separate 
parts whose invariants can be identified with $\sigma^2$ and $\omega^2$. Physically the shear corresponds to convergent flows, which will deform the halo, while the rotational part will give rise to an overall spinning of the halo induced by the external fields.}}
As already mentioned, angular momentum will only be sourced by the anti-symmetric part of the matrix product 
$\boldsymbol{X}$; the Hodge dual to the angular momentum is the tensor
\begin{equation}
 L_{ij} =  a^2 \dot D_+[\boldsymbol{I},\boldsymbol{\Psi}]_{ij}\;.
\end{equation}
Now, since the angular momentum can also be expressed as
\begin{equation}
 L_i = I_{ij} \omega_j\;,
\end{equation}
with angular velocity $\omega_j$, we can conclude that
\begin{equation}
 L_{ij} = I_{il}\omega_{lj}\;,
\end{equation}
and thus, in matrix-vector notation
\begin{equation}
 \boldsymbol{\omega} = \boldsymbol{I^{-1}} \boldsymbol{X}^{-}\;.
\end{equation}
For the shear we proceed in complete analogy, but using the anti-commutator instead of the commutator. In particular we 
decompose the tidal gravitational field as follows:
\begin{equation}\label{eq:decomp}
 \boldsymbol{\Psi} = \boldsymbol{I}^{-1}\boldsymbol{I}\boldsymbol{\Psi} = \boldsymbol{I}^{-1}\boldsymbol{X}^{+} + 
 \boldsymbol{I}^{-1}\boldsymbol{X}^{-} \equiv \boldsymbol{\tilde\sigma} +\boldsymbol{\tilde\omega}\;.
\end{equation}
Here we identified the shear tensor $\boldsymbol{\tilde\sigma}$ and the rotation tensor 
$\boldsymbol{\tilde\omega}$. {\rev{Since $\boldsymbol{\tilde\sigma}$ still carries a trace we need to subtract it to 
arrive at the following expressions for the shear tensor and rotation tensor respectively:}}
\begin{equation}\label{eq:invariants}
 \boldsymbol{\sigma} = \frac{1}{2}\left(\boldsymbol{\Psi} + \boldsymbol{I}^{-1}\boldsymbol{\Psi}\boldsymbol{I}\right) - 
                       \frac{\text{tr}\boldsymbol{\Psi}}{3}\mathbb{I}_3\;, \quad 
 \boldsymbol{\omega} = \frac{1}{2}\left(\boldsymbol{\Psi} - 
                       \boldsymbol{I}^{-1}\boldsymbol{\Psi}\boldsymbol{I}\right)\;.
\end{equation}
{\rev{The interpretation of the two expressions is straightforward: $\boldsymbol{\sigma}$ measures the alignment 
between the eigen-frames of the tidal tensor and the inertial tensor, while $\boldsymbol{\omega}$ measures their 
misalignment. Clearly, if both are completely aligned, the tidal tensor will not induce any rotation and only the shear 
effect is present. If, however, the two frames are not aligned the inertial tensor will start rotating into the frame 
of the tidal tensor and keep its rotation once the lever arm will reduce dramatically during collapse.}}

\subsection{Model comparison}
Having set up all the important relations, it is worthy to compare the models presented here with the one from R16 
and the phenomenological model in \citet{DelPopolo2013a,DelPopolo2013b}.

{\rev{The procedure of R16 is quite similar to the one outlined here. Values for the tidal tensor $\boldsymbol{\Psi}$ 
are sampled in the same way, the values for the inertial tensor, however, are not sampled from the density field and 
$\boldsymbol{I}$ is implicitly assumed to be the one of a spherical object. $\boldsymbol{I}$ is thus proportional to 
the identity, which itself commutes with every other tensor, thus setting $\boldsymbol{\omega}$ to zero identically. 
Especially this means Eq.~(\ref{eq:invariants}) was
\begin{equation}
 \boldsymbol{\sigma} = \boldsymbol{\Psi} - \frac{\mathrm{tr}\boldsymbol{\Psi}}{3}\mathbb{I}_3\;, \quad
                       \boldsymbol{\omega} = 0\;,
\end{equation}
for the model presented in R16, thus the inertial tensor is not needed as well as the condition to consider peaks in 
the density field only. 
This leads to a few subtle differences between the two models in terms of the physical interpretation: Both models 
describe the collapse of a spherically symmetric test object in a Gaussian random field. In both cases the tidal tensor 
$\boldsymbol{\Psi}$ is evaluated from the statistics of the underlying linearly evolved density field and gives rise to 
effective external forces which act on the collapse equation as an inhomogeneity. While the position of the test mass 
in R16 has been arbitrary, we restrict ourself to peaks in the density field here and include the possible spin up due 
to tidal torquing of the test mass. The restriction to peaks in the density field will generally lead to higher values 
in $\boldsymbol{\Psi}$ compared to R16, due to the non-trivial correlation with $\boldsymbol{I}$. In this sense the 
model presented here is more realistic, in terms of the shear and rotation being just inhomogeneities entering in the 
collapse equation, then the one in R16.}}

The comparison with \citet{DelPopolo2013b} is somewhat more difficult as their model was heuristically motivated 
only. In contrast our model relies on the statistics of the cosmic density field only and is in this sense only 
restricted by the validity of Lagrangian perturbation theory at first order. {\rev{This is certainly valid as long $\delta\ll 1$. If we are considering objects with masses above $10^14\; M_\odot$, this criterion is certainly satisfied in the sense of that the variance of the density field smoothed at this scale is well below unity.}}
In particular, \citet{DelPopolo2013b} find values for $\omega^2>\sigma^2$ which is not possible with our treatment. 
This is because in \citet{DelPopolo2013b}, the rotation term was derived to match the angular momentum of galaxies and 
clusters today, being therefore a non-linear quantity. This value will be exceeding our estimate of the rotation 
tensor and lead to effects that are opposite to what we find.

\subsection{Calculation of the invariant \texorpdfstring{$\sigma^2 - \omega^2$}{s2o2}}
The invariant quantities $\sigma^2$ and $\omega^2$ just differ by the sign of the cross terms and by the terms which 
arise due to the term including the trace of $\boldsymbol{\Psi}$. It is easy to see that the latter terms vanish 
identically, thus the only difference between $\sigma^2$ and $\omega^2$ is the sign of the two cross terms, which are 
themselves identical due to the cyclicity of the trace. Generalizing this reasoning to higher order invariants in a 
coordinate free way, we use that the invariants correspond to the Frobenius-norm of the tidal tensor and the inertia 
tensor. The Frobenius norm of a symmetric matrix $\boldsymbol{A}$ is defined as
\begin{equation}
 ||\boldsymbol{A}||^2 := \mathrm{tr}\boldsymbol{A}^2 \equiv A_{ij}A^{ji}\;.
\end{equation}
An inner product can be defined in the following way:
\begin{equation}
 \langle \boldsymbol{A},\boldsymbol{B}\rangle = \mathrm{tr}\boldsymbol{AB} = A_{ij}B^{ji}\;,
\end{equation}
which is also called Frobenius scalar product, then inducing the Frobenius norm defined above. With this we find
\begin{equation}\label{eq:positivedefinit}
 \begin{split}
  \sigma^2 = & \ ||\left\{\boldsymbol{I},\boldsymbol{\Psi}\right\}||^2 = ||\boldsymbol{I\Psi}||^2 + 2\langle 
                   \boldsymbol{I\Psi},\boldsymbol{\Psi I}\rangle + ||\boldsymbol{\Psi I}||^2 \\
  \omega^2 = & \ ||\left[\boldsymbol{I},\boldsymbol{\Psi}\right]||^2 = ||\boldsymbol{I\Psi}||^2 - 2\langle 
                   \boldsymbol{I\Psi},\boldsymbol{\Psi I}\rangle + ||\boldsymbol{\Psi I}||^2
 \end{split}
\end{equation}
Clearly, the positive definiteness of the Frobenius-norm implies that $\sigma^2>\omega^2$ is fulfilled if 
$\langle \boldsymbol{I\Psi},\boldsymbol{\Psi I}\rangle>0$. Due to the cyclic property of the trace this term can be 
shown to be $\langle\boldsymbol{I\Psi},\boldsymbol{\Psi I}\rangle = \mathrm{tr}(\boldsymbol{I\Psi}^2\boldsymbol{I}) = 
\mathrm{tr}(\boldsymbol{I}^2\boldsymbol{\Psi}^2) = \langle \boldsymbol{I}^2,\boldsymbol{\Psi}^2\rangle$, which in turn 
is positive for positive (semi-) definite matrices $\boldsymbol{I}$ and $\boldsymbol{\Psi}$.

To show this, one can use the generalisation of the inequality of the arithmetic and geometric mean,
\begin{equation}
 \frac{1}{n}\langle \boldsymbol{I}^2,\boldsymbol{\Psi}^2\rangle = 
 \frac{1}{n}\mathrm{tr}(\boldsymbol{I}^2\boldsymbol{\Psi}^2) \geq 
 \left(\mathrm{det}(\boldsymbol{I})\mathrm{det}(\boldsymbol{\Psi})\right)^\frac{2}{n}\geq 0\;,
 \label{eqn_trace_inequality}
\end{equation}
which is only valid for positive (semi-)definite matrices, with $n$ being the dimension of the matrices. The tidal 
shear is positive definite at a peak of the density field, because 
$\mathrm{tr}(\boldsymbol{\Psi}) = \Delta\Psi = \delta>0$ due to the Poisson-equation, and the inertia can only sensibly 
be defined at a maximum of the density field, where the curvature of the density field assumes positive values, 
resulting in a positive definite inertia $\mathrm{tr}(\boldsymbol{I})>0$. The argumentation applies for the traceless 
shear as well, as a positive semi-definite matrix. Both determinants are positive for positive definite matrices, and 
constrain the scalar product $\langle \boldsymbol{I}^2,\boldsymbol{\Psi}^2\rangle$ to be larger than zero. This is an 
important result, we find that the induced shear is always larger than the induced rotation by tidal torquing.

\autoref{Fig:sigmaomega} shows the effect mentioned in Eq.~(\ref{eq:positivedefinit}) very clearly: If we restrict the 
random process to maxima in the density field all values with $\sigma^2<\omega^2$ disappear and get shifted to larger 
ratios of $\sigma^2/\omega^2$. Thus, gravitational tidal fields will always introduce more shear than rotation by tidal 
torquing if only maxima of the underlying density field are considered. This is indeed a necessary condition, since 
otherwise the inertia tensor would not be defined in a proper way. As a consequence the collapse will always proceed 
faster in a scenario with tidal gravitational fields than in a uniform background as it is the case for the SPC.

We show the, not normalized, joint distribution of $\sigma^2$ and $\delta$ in \autoref{Fig:Correlations}. Due to the 
correlations in the $y^n_{lm}$ basis given in Eq.~(\ref{eq:correlations}), the maximum constraint enforces higher 
values in $\delta$ and $\sigma^2$. In particular, peaks can only be found if $\delta>0$, which is indeed necessary to 
write down the inertia tensor as in Eq.~(\ref{eq:inertialtensor}), as the ellipsoid is a region with boundary 
$\delta = 0$. Also the density peaks are significantly higher than without the constraint. 

In \autoref{Fig:Compareprevious} we show the distribution of $\sigma^2-\omega^2$ with different thresholds for the 
overdensity $\delta$ at the peak. Clearly, higher overdensities at the peak imply higher shear values as the potential 
is more curved at higher peaks.

\begin{figure}
 \begin{center}
  \includegraphics[width = 0.45\textwidth]{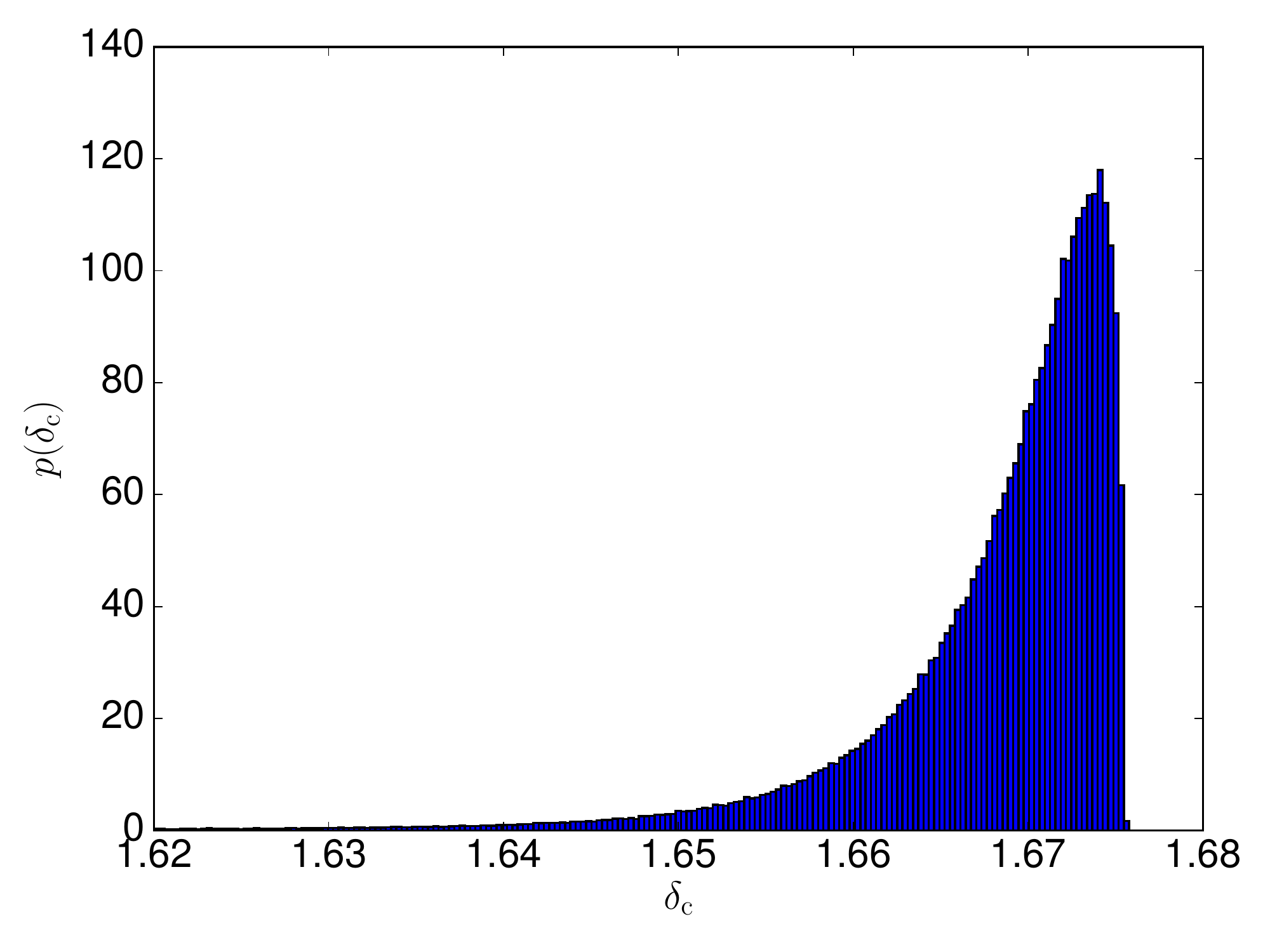}
  \caption{Distribution of the critical linear overdensity $\delta_\mathrm{c}$ for a standard $\Lambda$CDM cosmology. 
  The smoothing scale is again $R = 10 \text{Mpc}h^{-1}$ and the density threshold is $\delta=0$.}
  \label{Fig:DistDeltac}
 \end{center}
\end{figure}

\begin{figure*}
 \begin{center}
  \includegraphics[width = 0.45\textwidth]{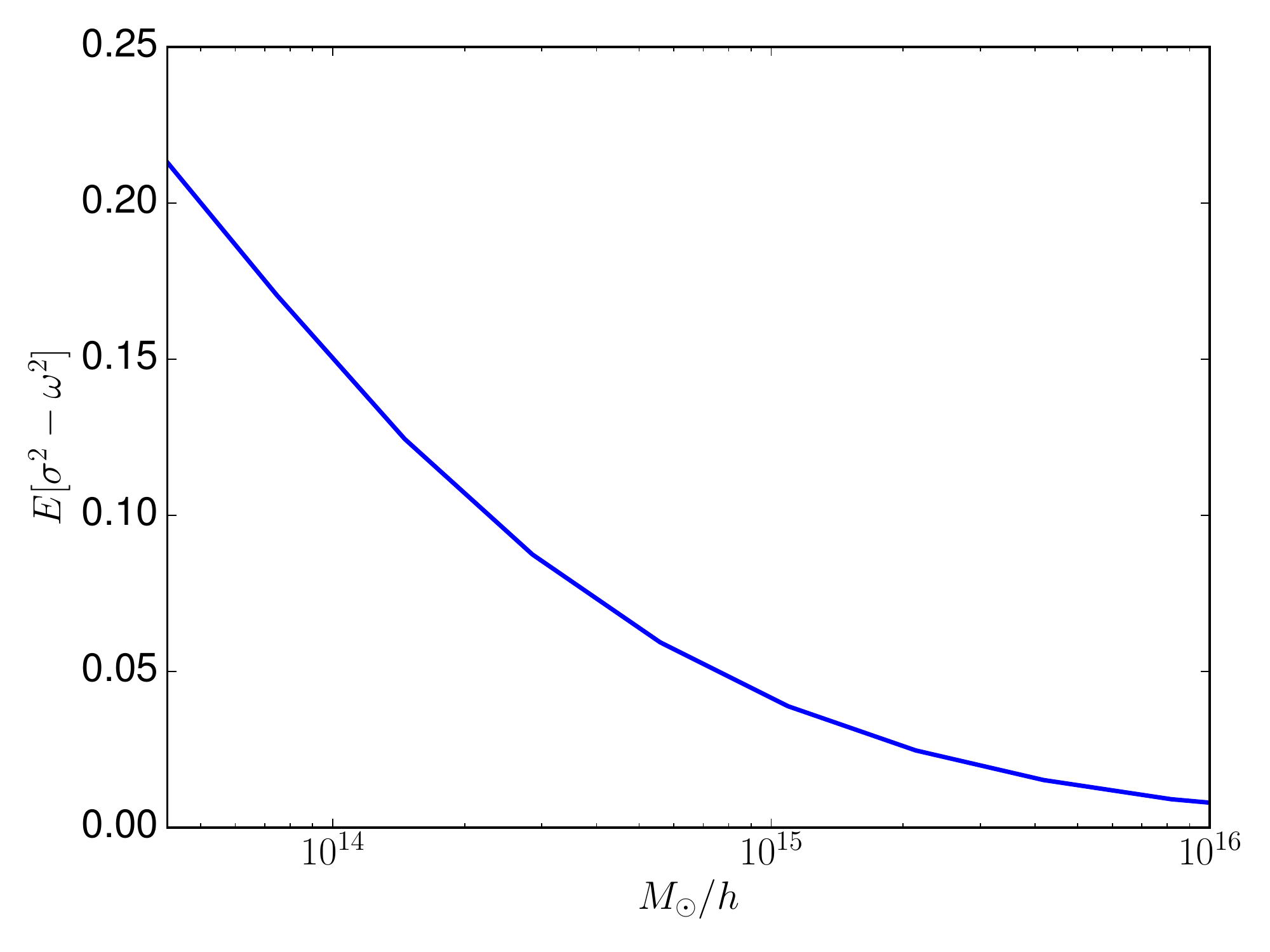}
  \includegraphics[width = 0.45\textwidth]{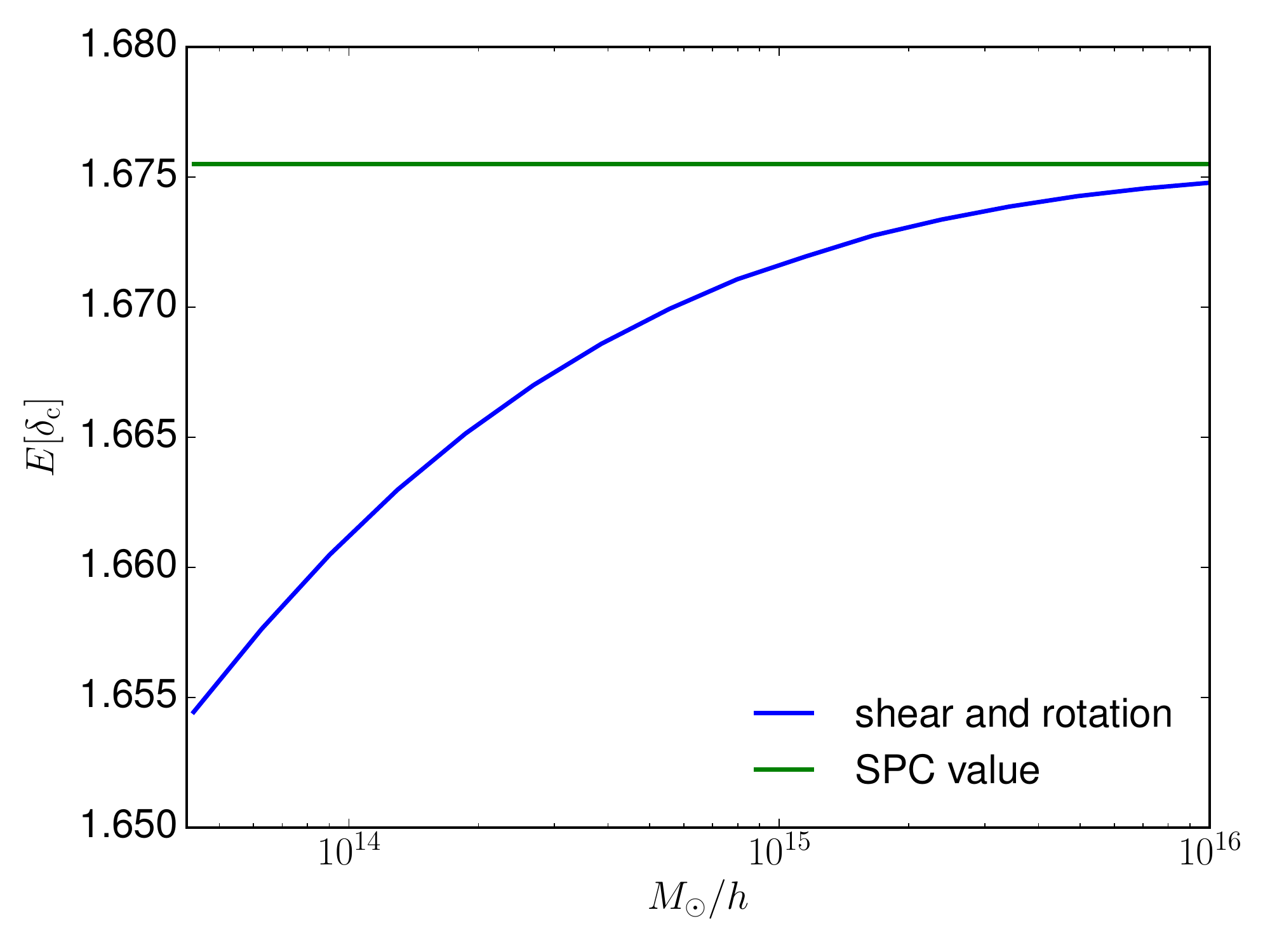}
  \caption{{\rev{Scaling relations of averaged quantities with mass evaluated at redshift zero. The blue curve shows 
  the effect derived in this paper by including both $\sigma^2$ and $\omega^2$. The influence of gravitational tidal 
  fields is highest for low masses, thus $\delta_\mathrm{c}$ is also influenced most at the low-mass tail. 
  \textit{Left:} Averaged invariants, \textit{Right:} averaged $\delta_\mathrm{c}$. The value for the SPC for a 
  $\Lambda$CDM universe is shown in green.}}}
  \label{Fig:Scaling}
 \end{center}
\end{figure*}

\section{Influence on \texorpdfstring{$\delta_{\rm c}$}{dc}, \texorpdfstring{$\Delta_{\rm V}$}{Dv} and Scaling 
Properties}\label{sec:deltac}
In this section we investigate the influence of the tidal gravitational fields on the collapse dynamics by substituting 
the invariants $\sigma^2$ and $\omega^2$ into the collapse equation. Additionally we will study the scaling with the 
mass of the collapsed structure. The cosmology is chosen to be a concordance $\Lambda$CDM model with 
$\Omega_\mathrm{m} = 0.3$, $\Omega_\Lambda = 0.7$, $w = -1$, $h=0.7$, $\sigma_8 = 0.8$ and $n_\mathrm{s} = 0.96$.

In \autoref{Fig:DistDeltac} the resulting distribution of $\delta_\mathrm{c}$ is shown. The collapse always proceeds 
faster than in the case without tidal fields. {\rev{For more work on this we refer to 
\citet{Hoffman1986,Zaroubi1993,Bertschinger1994a}}}. 
As discussed in the previous section, this is due to the fact that the tidal field induced shear is always higher than 
the effect due to tidal torquing, provided we restrict our considerations to maxima in the density field. Thus the 
strong drop of the distribution at higher $\delta_\mathrm{c}$ marks the value which one would get within a uniform 
background.

Due to the faster collapse, virialised objects form more easily, thus yielding more massive objects. This effect is 
similar to modified gravities theories or dark energy cosmologies with non-phantom equations of state. Since the 
distribution found for $\delta_\mathrm{c}$ is similar to the one found in R16 and no significant differences were 
found for more complex dark energy models, we refer to our previous works regarding the impact on the mass function and 
cluster counts \citep[see][]{Reischke2016a,Pace2017}.

$\delta_\mathrm{c}$ exhibits a mass dependence due to the low-pass filter with a scale $R$ which is 
introduced to model the effective tidal fields acting on an object of size $R(M)$. 
We consider again the averaged values of the invariant $\sigma^2-\omega^2$ or the linear critical density contrast 
$\delta_\mathrm{c}${\rev{, i.e. given the distribution $p(\delta_\mathrm{c})$ we consider
\begin{equation}
 E[\delta_\mathrm{c}] = \int p(\delta_\mathrm{c})\delta_\mathrm{c}\mathrm{d}\delta_\mathrm{c}\;,
\end{equation}
and similarly for $\sigma^2-\omega^2$.}} 
{\rev{On the left panel in \autoref{Fig:Scaling} we show the scaling of $E[\sigma^2-\omega^2]$ with respect to the 
mass. 
The general scaling shows that higher masses result into lower values for $\sigma^2-\omega^2$ as larger objects are 
only influenced by low frequency modes which become smaller for increasing scale. 
In the case considered here we restrict ourselves to maxima in the density field, thus the situation is constructed 
such that the curvature of the density field must be negative, yielding slightly more shear on large scales than for a 
random point in the density field. On smaller scales, however, the situation is reversed. 
This argument is precisely due to the additional factor $k^4$ which enters in the random process for $\delta_{ij}$ 
(cf. Eqs.~\ref{eq:spectralmoments} and \ref{eq:curvaturedens}).}}

{\rev{The right panel of \autoref{Fig:Scaling} shows the resulting scaling of $E[\delta_\mathrm{c}]$. Here we 
additionally show the constant value (green curve) obtained without gravitational tidal fields. As for the invariants 
$\sigma^2$ and $\omega^2$ the qualitative behaviour is identical. We find that the the term $\sigma^2-\omega^2$ will 
always favour the collapse, thus lowering $E[\delta_c]$. Even though $\omega^2$ will act against the collapse, as it 
corresponds to a centrifugal force, it can never dominate $\sigma^2$ as we showed before. For completeness we note that 
our final results for $E[\delta_\mathrm{c}]$ are effectively very similar to the ones found in R16. 
Furthermore we note that the time evolution of the invariant is controlled by the time derivative of the growth factor 
introduced in Eq.~(\ref{eq:Zeld}) and is thus purely due to background dynamics. If one instead starts with a non-spherical collapse, one would find larger effects on $\delta_\mathrm{c}$ compared to this idealised model. An example for this is the ellipsoidal collapse model \citep{Eisenstein1995,Ohta2003,Ohta2004,Angrick2010}, where $\delta_\mathrm{c}$ values are normally substantially higher than for the spherical collapse case, especially at low redshift and mass.}}

A very important and interesting quantity that can be evaluated within the framework of the spherical collapse model is 
the virial overdensity $\Delta_{\rm V}$, representing the overdensity of the collapsing object at the virialisation 
epoch \citep[see also][for a discussion of this quantity in a general relativistic setting]{Meyer2012}. The virial 
overdensity is also related to the size of spherically symmetric halos and its value can be inferred by embedding the 
virial theorem into the formalism. 
{\rev{When including the shear and the rotation terms into the equations of motion for dark matter perturbations, 
$\Delta_{\rm V}$ becomes, in analogy to $\delta_{\rm c}$, mass-dependent. However, one finds that $\Delta_{\rm V}$ is 
practically independent of mass and it evolves as if the system is evolving in a ideal background, i.e. without shear 
and rotation. This is an interesting result but not unexpected. As showed in R16, the virial overdensity is insensitive 
to mass since the quantities involved for its determination are evaluated still in the linear regime and perturbations 
with respect to the spherically symmetric case are of the order of per mill. Taking also into account that rotation has 
always a smaller contribution than the shear and their combined effect makes the rotating ellipsoid closer to the 
sphere in terms of the perturbation quantities, it is easy to understand why the feature found in R16 still holds.}}

\section{Conclusion and Discussion}\label{sec:concl}
{\rev{In this paper we extended the work by R16 to estimate the effect of shear and rotation on the spherical collapse 
of dark matter halos. The model assumes that the spherical collapse dynamics are only altered in terms of a 
inhomogeneity in the collapse equation which also only enters in the non-linear equation. In this sense the model 
describes the spherical collapse of a test mass in a tidal gravitational field.}}

By jointly considering the gravitational tidal field and the curvature of the density field we separated its action 
into a symmetric traceless part and an anti-symmetric part which correspond to the shear tensor and rotation tensor 
respectively. These tensors were used to construct the invariants $\sigma^2$ and $\omega^2$ in the collapse equation. 
Physically, the protohalo, forming at the location of a peak, feels the surrounding tidal gravitational field and thus 
shear effects as well as rotation induced by tidal torquing. 
{\rev{This procedure is identical to the one presented in R16 if we restrict our considerations again to peaks with 
spherical symmetry. Our findings are the following:}}

\begin{enumerate}
\item
The invariant quantity $\omega^2$ of the rotational part of the tidal tensor is always smaller than the shear invariant 
$\sigma^2$ within the framework of tidal torquing. This statement is not of statistical nature, it is true for every 
sample individually.

\item {\rev{The critical linear overdensity $\delta_\mathrm{c}$ is now a mass dependent quantity changing by roughly a 
percent with respect to the usual spherical collapse value. The overall effect is small at masses below 
$10^{15}\; M_{\odot}$ and completely negligible for masses above.}}

\item {\rev{External tidal fields will always help objects to collapse into virialised structures even if a rotational 
term due to tidal torquing is considered. In terms of observations of cluster counts tidal fields can in principle 
always be confused with dynamical dark energy increasing the abundance of heavy clusters in a purely spherically 
symmetric case where no tidal fields are taken into account. For a more detailed discussion on this, we refer the 
reader to R16.}}

\item
Comparing this work with \citet{DelPopolo2013a} we find that the deviations of $\delta_\mathrm{c}$ found there are 
mainly due to the rotational term, which can become rather large, thus the collapse is mostly slowed down. 
Our work finds an opposite result as the gravitational tidal fields always speed up the collapse and the rotational 
term is nearly negligible. This is, however, also a property of the model we used here. Our model is self-consistent in 
as long as we only consider external tidal effects on a spherically symmetric object where the deformation is 
negligible compared to the total extent of the collapsing object. In this work we assumed the halo to be non-spherical 
prior to collapse to allow it to spin up as long as the lever arms are large enough. As soon as collapse starts, the 
collapse is again treated as being spherical. We therefore have a situation in which a spherical overdensity is 
rotating at an angular speed $\omega^2$ gained by tidal torquing as if it would have been an ellipsoidal object. These 
limitations make a direct comparison with \citet{DelPopolo2013a} difficult. See point \ref{itm:comp} for an explanation 
based on the way the invariant $\sigma^2-\omega^2$ is evaluated.

\item \label{itm:comp}
In our self-consistent model, the shear and rotation term have little effect and their effect grows with time and 
mass as structures evolve. In the formalism outlined, the invariant $\sigma^2-\omega^2$ is evaluated at early times 
when structures are in the linear regime. This explains why, for example, the virial overdensity $\Delta_{\rm V}$ is 
barely affected. In previous works on the subject \citep{DelPopolo2013a,DelPopolo2013b,Pace2014b} instead, the term 
$\sigma^2-\omega^2$ assumes objects to be still spherical in average and that the rotation term matches the present-day 
rotational velocity of clusters as a function of their mass. This late time evaluation makes the rotation term 
$\omega^2$ the dominant one and this explains the different trends in the two series of papers.

\end{enumerate}

\section{Acknowledgements}
RR acknowledges funding by the graduate college "Astrophysics of cosmological probes of gravity" by 
Landesgraduiertenakademie Baden-W\"urttemberg. 
FP is supported by the STFC post-doctoral fellowship with grant R120562 'Astrophysics and Cosmology Research within the 
JBCA 2017-2020' and thanks Inga Cebotaru for reading the manuscript and providing useful comments. The authors thank an 
anonymous referee for improving the manuscript.

\bibliographystyle{mnras}
\bibliography{../../Bib/MyBiB,../../Bib/old_MasterBib}

\begin{thebibliography}{}
\makeatletter
\relax
\def\mn@urlcharsother{\let\do\@makeother \do\$\do\&\do\#\do\^\do\_\do\%\do\~}
\def\mn@doi{\begingroup\mn@urlcharsother \@ifnextchar [ {\mn@doi@}
  {\mn@doi@[]}}
\def\mn@doi@[#1]#2{\def\@tempa{#1}\ifx\@tempa\@empty \href
  {http://dx.doi.org/#2} {doi:#2}\else \href {http://dx.doi.org/#2} {#1}\fi
  \endgroup}
\def\mn@eprint#1#2{\mn@eprint@#1:#2::\@nil}
\def\mn@eprint@arXiv#1{\href {http://arxiv.org/abs/#1} {{\tt arXiv:#1}}}
\def\mn@eprint@dblp#1{\href {http://dblp.uni-trier.de/rec/bibtex/#1.xml}
  {dblp:#1}}
\def\mn@eprint@#1:#2:#3:#4\@nil{\def\@tempa {#1}\def\@tempb {#2}\def\@tempc
  {#3}\ifx \@tempc \@empty \let \@tempc \@tempb \let \@tempb \@tempa \fi \ifx
  \@tempb \@empty \def\@tempb {arXiv}\fi \@ifundefined
  {mn@eprint@\@tempb}{\@tempb:\@tempc}{\expandafter \expandafter \csname
  mn@eprint@\@tempb\endcsname \expandafter{\@tempc}}}

\bibitem[\protect\citeauthoryear{{Abramo}, {Batista}, {Liberato}  \&
  {Rosenfeld}}{{Abramo} et~al.}{2007}]{Abramo2007}
{Abramo} L.~R.,  {Batista} R.~C.,  {Liberato} L.,   {Rosenfeld} R.,  2007,
  \mn@doi [Journal of Cosmology and Astro-Particle Physics]
  {10.1088/1475-7516/2007/11/012}, \href
  {http://adsabs.harvard.edu/abs/2007JCAP...11..012A} {11, 12}

\bibitem[\protect\citeauthoryear{{Abramo}, {Batista}  \& {Rosenfeld}}{{Abramo}
  et~al.}{2009}]{Abramo2009a}
{Abramo} L.~R.,  {Batista} R.~C.,   {Rosenfeld} R.,  2009, \mn@doi [Journal of
  Cosmology and Astro-Particle Physics] {10.1088/1475-7516/2009/07/040}, \href
  {http://adsabs.harvard.edu/abs/2009JCAP...07..040A} {7, 40}

\bibitem[\protect\citeauthoryear{Angrick \& Bartelmann}{Angrick \&
  Bartelmann}{2009}]{Angrick2009}
Angrick C.,  Bartelmann M.,  2009, Astronomy \& Astrophysics, 494, 461

\bibitem[\protect\citeauthoryear{Angrick \& Bartelmann}{Angrick \&
  Bartelmann}{2010}]{Angrick2010}
Angrick C.,  Bartelmann M.,  2010, Astronomy \& Astrophysics, 518, A38

\bibitem[\protect\citeauthoryear{{Avila-Reese}, {Firmani}  \&
  {Hern{\'a}ndez}}{{Avila-Reese} et~al.}{1998}]{AvilaReese1998}
{Avila-Reese} V.,  {Firmani} C.,   {Hern{\'a}ndez} X.,  1998, \mn@doi [\apj]
  {10.1086/306136}, \href {http://adsabs.harvard.edu/abs/1998ApJ...505...37A}
  {505, 37}

\bibitem[\protect\citeauthoryear{{Bernardeau}}{{Bernardeau}}{1994}]{Bernardeau1994}
{Bernardeau} F.,  1994, \mn@doi [\apj] {10.1086/174620}, \href
  {http://adsabs.harvard.edu/abs/1994ApJ...433....1B} {433, 1}

\bibitem[\protect\citeauthoryear{{Bertschinger}}{{Bertschinger}}{1985}]{Bertschinger1985}
{Bertschinger} E.,  1985, \mn@doi [\apjs] {10.1086/191028}, \href
  {http://adsabs.harvard.edu/abs/1985ApJS...58...39B} {58, 39}

\bibitem[\protect\citeauthoryear{{Bertschinger} \& {Jain}}{{Bertschinger} \&
  {Jain}}{1994}]{Bertschinger1994a}
{Bertschinger} E.,  {Jain} B.,  1994, \mn@doi [\apj] {10.1086/174501}, \href
  {http://adsabs.harvard.edu/abs/1994ApJ...431..486B} {431, 486}

\bibitem[\protect\citeauthoryear{{Catelan} \& {Theuns}}{{Catelan} \&
  {Theuns}}{1996}]{Catelan1996a}
{Catelan} P.,  {Theuns} T.,  1996, \mnras, \href
  {http://adsabs.harvard.edu/abs/1996MNRAS.282..436C} {282, 436}

\bibitem[\protect\citeauthoryear{{Cole} et~al.,}{{Cole}
  et~al.}{2005}]{Cole2005}
{Cole} S.,  et~al., 2005, \mn@doi [\mnras] {10.1111/j.1365-2966.2005.09318.x},
  \href {http://adsabs.harvard.edu/abs/2005MNRAS.362..505C} {362, 505}

\bibitem[\protect\citeauthoryear{{Copeland}, {Sami}  \& {Tsujikawa}}{{Copeland}
  et~al.}{2006}]{Copeland2006}
{Copeland} E.~J.,  {Sami} M.,   {Tsujikawa} S.,  2006, \mn@doi [International
  Journal of Modern Physics D] {10.1142/S021827180600942X}, \href
  {http://adsabs.harvard.edu/abs/2006IJMPD..15.1753C} {15, 1753}

\bibitem[\protect\citeauthoryear{{Crittenden}, {Natarajan}, {Pen}  \&
  {Theuns}}{{Crittenden} et~al.}{2001}]{Crittenden2001}
{Crittenden} R.~G.,  {Natarajan} P.,  {Pen} U.-L.,   {Theuns} T.,  2001,
  \mn@doi [\apj] {10.1086/322370}, \href
  {http://adsabs.harvard.edu/abs/2001ApJ...559..552C} {559, 552}

\bibitem[\protect\citeauthoryear{{Del Popolo}, {Pace}  \& {Lima}}{{Del Popolo}
  et~al.}{2013a}]{DelPopolo2013a}
{Del Popolo} A.,  {Pace} F.,   {Lima} J.~A.~S.,  2013a, \mn@doi [International
  Journal of Modern Physics D] {10.1142/S0218271813500387}, \href
  {http://adsabs.harvard.edu/abs/2013IJMPD..2250038D} {22, 50038}

\bibitem[\protect\citeauthoryear{{Del Popolo}, {Pace}  \& {Lima}}{{Del Popolo}
  et~al.}{2013b}]{DelPopolo2013b}
{Del Popolo} A.,  {Pace} F.,   {Lima} J.~A.~S.,  2013b, \mn@doi [\mnras]
  {10.1093/mnras/sts669}, \href
  {http://adsabs.harvard.edu/abs/2013MNRAS.430..628D} {430, 628}

\bibitem[\protect\citeauthoryear{{Diego} \& {Majumdar}}{{Diego} \&
  {Majumdar}}{2004}]{Diego2004}
{Diego} J.~M.,  {Majumdar} S.,  2004, \mn@doi [\mnras]
  {10.1111/j.1365-2966.2004.07989.x}, \href
  {http://adsabs.harvard.edu/abs/2004MNRAS.352..993D} {352, 993}

\bibitem[\protect\citeauthoryear{{Eisenstein} \& {Loeb}}{{Eisenstein} \&
  {Loeb}}{1995}]{Eisenstein1995}
{Eisenstein} D.~J.,  {Loeb} A.,  1995, \mn@doi [\apj] {10.1086/175193}, \href
  {http://adsabs.harvard.edu/abs/1995ApJ...439..520E} {439, 520}

\bibitem[\protect\citeauthoryear{{Fang} \& {Haiman}}{{Fang} \&
  {Haiman}}{2007}]{Fang2007}
{Fang} W.,  {Haiman} Z.,  2007, \mn@doi [\prd] {10.1103/PhysRevD.75.043010},
  \href {http://adsabs.harvard.edu/abs/2007PhRvD..75d3010F} {75, 043010}

\bibitem[\protect\citeauthoryear{{Fillmore} \& {Goldreich}}{{Fillmore} \&
  {Goldreich}}{1984}]{Fillmore1984}
{Fillmore} J.~A.,  {Goldreich} P.,  1984, \mn@doi [\apj] {10.1086/162070},
  \href {http://adsabs.harvard.edu/abs/1984ApJ...281....1F} {281, 1}

\bibitem[\protect\citeauthoryear{{Gunn} \& {Gott}}{{Gunn} \&
  {Gott}}{1972}]{Gunn1972}
{Gunn} J.~E.,  {Gott} III J.~R.,  1972, \mn@doi [\apj] {10.1086/151605}, \href
  {http://adsabs.harvard.edu/abs/1972ApJ...176....1G} {176, 1}

\bibitem[\protect\citeauthoryear{Heavens \& Sheth}{Heavens \&
  Sheth}{1999}]{Heavens1999}
Heavens A.~F.,  Sheth R.~K.,  1999, Monthly Notices of the Royal Astronomical
  Society, 310, 1062

\bibitem[\protect\citeauthoryear{{Hoffman}}{{Hoffman}}{1986}]{Hoffman1986}
{Hoffman} Y.,  1986, \mn@doi [\apj] {10.1086/164520}, \href
  {http://adsabs.harvard.edu/abs/1986ApJ...308..493H} {308, 493}

\bibitem[\protect\citeauthoryear{{Komatsu}, {Smith}, {Dunkley}  \&
  {et~al.}}{{Komatsu} et~al.}{2011}]{Komatsu2011}
{Komatsu} E.,  {Smith} K.~M.,  {Dunkley} J.,   {et~al.} 2011, \mn@doi [\apjs]
  {10.1088/0067-0049/192/2/18}, \href
  {http://adsabs.harvard.edu/abs/2011ApJS..192...18K} {192, 18}

\bibitem[\protect\citeauthoryear{Lin \& Kilbinger}{Lin \&
  Kilbinger}{2014}]{Lin2014}
Lin C.-A.,  Kilbinger M.,  2014, Proceedings of the International Astronomical
  Union, 10, 107

\bibitem[\protect\citeauthoryear{{Majumdar}}{{Majumdar}}{2004}]{Majumdar2004}
{Majumdar} S.,  2004, \mn@doi [Pramana] {10.1007/BF02705209}, \href
  {http://adsabs.harvard.edu/abs/2004Prama..63..871M} {63, 871}

\bibitem[\protect\citeauthoryear{{Maturi}, {Angrick}, {Pace}  \&
  {Bartelmann}}{{Maturi} et~al.}{2010}]{Maturi2010}
{Maturi} M.,  {Angrick} C.,  {Pace} F.,   {Bartelmann} M.,  2010, \mn@doi
  [\aap] {10.1051/0004-6361/200912866}, \href
  {http://adsabs.harvard.edu/abs/2010A%26A...519A..23M} {519, A23}

\bibitem[\protect\citeauthoryear{Maturi, Fedeli  \& Moscardini}{Maturi
  et~al.}{2011}]{Maturi2011}
Maturi M.,  Fedeli C.,   Moscardini L.,  2011, Monthly Notices of the Royal
  Astronomical Society, 416, 2527

\bibitem[\protect\citeauthoryear{{Meyer}, {Pace}  \& {Bartelmann}}{{Meyer}
  et~al.}{2012}]{Meyer2012}
{Meyer} S.,  {Pace} F.,   {Bartelmann} M.,  2012, \mn@doi [\prd]
  {10.1103/PhysRevD.86.103002}, \href
  {http://adsabs.harvard.edu/abs/2012PhRvD..86j3002M} {86, 103002}

\bibitem[\protect\citeauthoryear{{Mota} \& {van de Bruck}}{{Mota} \& {van de
  Bruck}}{2004}]{Mota2004}
{Mota} D.~F.,  {van de Bruck} C.,  2004, \mn@doi [\aap]
  {10.1051/0004-6361:20041090}, \href
  {http://adsabs.harvard.edu/abs/2004A%26A...421...71M} {421, 71}

\bibitem[\protect\citeauthoryear{{Ohta}, {Kayo}  \& {Taruya}}{{Ohta}
  et~al.}{2003}]{Ohta2003}
{Ohta} Y.,  {Kayo} I.,   {Taruya} A.,  2003, \mn@doi [\apj] {10.1086/374375},
  \href {http://adsabs.harvard.edu/abs/2003ApJ...589....1O} {589, 1}

\bibitem[\protect\citeauthoryear{{Ohta}, {Kayo}  \& {Taruya}}{{Ohta}
  et~al.}{2004}]{Ohta2004}
{Ohta} Y.,  {Kayo} I.,   {Taruya} A.,  2004, \mn@doi [\apj] {10.1086/420762},
  \href {http://adsabs.harvard.edu/abs/2004ApJ...608..647O} {608, 647}

\bibitem[\protect\citeauthoryear{{Pace}, {Waizmann}  \& {Bartelmann}}{{Pace}
  et~al.}{2010}]{Pace2010}
{Pace} F.,  {Waizmann} J.-C.,   {Bartelmann} M.,  2010, \mn@doi [\mnras]
  {10.1111/j.1365-2966.2010.16841.x}, \href
  {http://adsabs.harvard.edu/abs/2010MNRAS.406.1865P} {406, 1865}

\bibitem[\protect\citeauthoryear{{Pace}, {Moscardini}, {Crittenden},
  {Bartelmann}  \& {Pettorino}}{{Pace} et~al.}{2014a}]{Pace2014}
{Pace} F.,  {Moscardini} L.,  {Crittenden} R.,  {Bartelmann} M.,   {Pettorino}
  V.,  2014a, \mn@doi [\mnras] {10.1093/mnras/stt1907}, \href
  {http://adsabs.harvard.edu/abs/2014MNRAS.437..547P} {437, 547}

\bibitem[\protect\citeauthoryear{{Pace}, {Batista}  \& {Del Popolo}}{{Pace}
  et~al.}{2014b}]{Pace2014b}
{Pace} F.,  {Batista} R.~C.,   {Del Popolo} A.,  2014b, \mn@doi [\mnras]
  {10.1093/mnras/stu1782}, \href
  {http://adsabs.harvard.edu/abs/2014MNRAS.445..648P} {445, 648}

\bibitem[\protect\citeauthoryear{{Pace}, {Reischke}, {Meyer}  \&
  {Sch{\"a}fer}}{{Pace} et~al.}{2017}]{Pace2017}
{Pace} F.,  {Reischke} R.,  {Meyer} S.,   {Sch{\"a}fer} B.~M.,  2017, \mn@doi
  [\mnras] {10.1093/mnras/stw3244}, \href
  {http://adsabs.harvard.edu/abs/2017MNRAS.466.1839P} {466, 1839}

\bibitem[\protect\citeauthoryear{{Padmanabhan}}{{Padmanabhan}}{1996}]{Padmanabhan1996}
{Padmanabhan} T.,  1996, {Cosmology and Astrophysics through Problems}

\bibitem[\protect\citeauthoryear{{Perlmutter}, {Aldering}, {Goldhaber}  \& {et
  al.}}{{Perlmutter} et~al.}{1999}]{Perlmutter1999}
{Perlmutter} S.,  {Aldering} G.,  {Goldhaber} G.,   {et al.} 1999, \mn@doi
  [\apj] {10.1086/307221}, \href
  {http://adsabs.harvard.edu/abs/1999ApJ...517..565P} {517, 565}

\bibitem[\protect\citeauthoryear{{Planck Collaboration} et~al.,}{{Planck
  Collaboration} et~al.}{2016}]{Planck2016_XIII}
{Planck Collaboration} et~al., 2016, \mn@doi [\aap]
  {10.1051/0004-6361/201525830}, \href
  {http://adsabs.harvard.edu/abs/2016A\%26A...594A..13P} {594, A13}

\bibitem[\protect\citeauthoryear{Reg\H{o}s \& Szalay}{Reg\H{o}s \&
  Szalay}{1995}]{Regos1995}
Reg\H{o}s E.,  Szalay A.~S.,  1995, Monthly Notices of the Royal Astronomical
  Society, 272, 447

\bibitem[\protect\citeauthoryear{{Reischke}, {Pace}, {Meyer}  \&
  {Sch{\"a}fer}}{{Reischke} et~al.}{2016a}]{Reischke2016a}
{Reischke} R.,  {Pace} F.,  {Meyer} S.,   {Sch{\"a}fer} B.~M.,  2016a, \mn@doi
  [\mnras] {10.1093/mnras/stw1989}, \href
  {http://adsabs.harvard.edu/abs/2016MNRAS.tmp.1091R} {}

\bibitem[\protect\citeauthoryear{Reischke, Maturi  \& Bartelmann}{Reischke
  et~al.}{2016b}]{Reischke2016}
Reischke R.,  Maturi M.,   Bartelmann M.,  2016b, Monthly Notices of the Royal
  Astronomical Society, 456, 641

\bibitem[\protect\citeauthoryear{{Riess}, {Filippenko}, {Challis}  \& {et
  al.}}{{Riess} et~al.}{1998}]{Riess1998}
{Riess} A.~G.,  {Filippenko} A.~V.,  {Challis} P.,   {et al.} 1998, \mn@doi
  [\aj] {10.1086/300499}, \href
  {http://adsabs.harvard.edu/abs/1998AJ....116.1009R} {116, 1009}

\bibitem[\protect\citeauthoryear{{Ryden} \& {Gunn}}{{Ryden} \&
  {Gunn}}{1987}]{Ryden1987}
{Ryden} B.~S.,  {Gunn} J.~E.,  1987, \mn@doi [\apj] {10.1086/165349}, \href
  {http://adsabs.harvard.edu/abs/1987ApJ...318...15R} {318, 15}

\bibitem[\protect\citeauthoryear{{Sch{\"a}fer}}{{Sch{\"a}fer}}{2009}]{Schafer2009}
{Sch{\"a}fer} B.~M.,  2009, \mn@doi [International Journal of Modern Physics D]
  {10.1142/S0218271809014388}, \href
  {http://adsabs.harvard.edu/abs/2009IJMPD..18..173S} {18, 173}

\bibitem[\protect\citeauthoryear{Sch{\"a}fer \& Koyama}{Sch{\"a}fer \&
  Koyama}{2008}]{Schafer2008}
Sch{\"a}fer B.~M.,  Koyama K.,  2008, Monthly Notices of the Royal Astronomical
  Society, 385, 411

\bibitem[\protect\citeauthoryear{Sch{\"a}fer \& Merkel}{Sch{\"a}fer \&
  Merkel}{2012}]{Schafer2012}
Sch{\"a}fer B.~M.,  Merkel P.~M.,  2012, Monthly Notices of the Royal
  Astronomical Society, 421, 2751

\bibitem[\protect\citeauthoryear{{Sunyaev} \& {Zeldovich}}{{Sunyaev} \&
  {Zeldovich}}{1980}]{Sunyaev1980b}
{Sunyaev} R.~A.,  {Zeldovich} I.~B.,  1980, \mn@doi [\araa]
  {10.1146/annurev.aa.18.090180.002541}, \href
  {http://adsabs.harvard.edu/abs/1980ARA%26A..18..537S} {18, 537}

\bibitem[\protect\citeauthoryear{{White}}{{White}}{1984}]{White1984}
{White} S.~D.~M.,  1984, \mn@doi [\apj] {10.1086/162573}, \href
  {http://adsabs.harvard.edu/abs/1984ApJ...286...38W} {286, 38}

\bibitem[\protect\citeauthoryear{{Zaroubi} \& {Hoffman}}{{Zaroubi} \&
  {Hoffman}}{1993}]{Zaroubi1993}
{Zaroubi} S.,  {Hoffman} Y.,  1993, \mn@doi [\apj] {10.1086/173246}, \href
  {http://adsabs.harvard.edu/abs/1993ApJ...416..410Z} {416, 410}

\bibitem[\protect\citeauthoryear{{Zel'Dovich}}{{Zel'Dovich}}{1970}]{zeldovich1970}
{Zel'Dovich} Y.~B.,  1970, \aap, \href
  {http://adsabs.harvard.edu/abs/1970A%26A.....5...84Z} {5, 84}

\makeatother
\end{thebibliography}

\label{lastpage}

\end{document}